\documentclass[journal, twoside, onecolumn, draftclsnofoot,12pt]{IEEEtran}

\usepackage{graphicx}
\DeclareGraphicsExtensions{.pdf,.jpg}

\ifCLASSINFOpdf
\else
\fi
\usepackage{amsmath}
\begin{document}
%
\title{Designing a Robust Carrier Frequency Offset Estimation Scheme for Meeting Target Decoding Performance in an OFDM System}
%
%
%

\author{Minkyeong Jeong~\IEEEmembership{Student Member,~IEEE},
        ~Sang-Won~Choi,~\IEEEmembership{Member,~IEEE}
        ~and~Juyeop~Kim,~\IEEEmembership{Member,~IEEE}
\thanks{ M. Jeong is with Sookmyung Women's University, Republic of Korea, email: ujk767@sookmyung.ac.kr.}
\thanks{ S.-W. Choi is with Kyonggi University, Republic of Korea, email: swchoi@kyonggi.ac.kr.}
\thanks{J. Kim is with Sookmyung Women's University, Republic of Korea, email: jykim@sookmyung.ac.kr.}
}
\maketitle

\begin{abstract}
Carrier Frequency Offset(CFO) is critical to the decoder performance in an Orthogonal Frequency Division Multiplexing(OFDM) system since phase distortions of received symbols becomes worse as time goes by.
A number of conventional CFO estimators have different characteristics in the perspectives of accuracy and estimation range and a proper estimator should be adopted in order for the receiver to have the satisfactory decoder performance.
In this paper, we propose a numerical model for designing an overall CFO estimation scheme for a specific OFDM system that has target decoding performance.
The numerical model generalizes CFO estimators that measures phase difference at two distinct moments and evaluates how robust a CFO estimator can get CFO in a noisy environment.
Based on our numerical model, we design a two-step CFO estimation scheme which is suitable to a Long Term Evolution(LTE) system. 
The proposed CFO estimation scheme has coarse and residual CFO estimators with different phase-measurement time interval so that it can estimate CFO accurately and in a wide range.
In addition, we implement the proposed CFO estimation scheme on a real-time operating testbed and evaluate the decoder performance of LTE Physical Broadcast CHannel(PBCH).
The experiment results show that the coarse and residual estimators implemented on the OpenAirInterface source code co-work well in various Signal to Noise Ratio(SNR) and frequency offset environments.

\end{abstract}


%
\IEEEpeerreviewmaketitle

\section{Introduction}
%
%
%
%

\IEEEPARstart{C}{urrent} communication systems have been evolved with adopting new technologies that can innovatively enhance their performance.
Meanwhile, the issue of Carrier Frequency Offset(CFO) still matters in these current communication systems in view of achieving the goal of performance enhancement in a lower Signal-to-Noise Ratio(SNR) environment.
Especially, estimating precise CFO has been dealt as a difficult problem from the earlier mobile communication systems, and is critical to the performance of Long Term Evolution(LTE) or 5G receivers.
The frequency offset happens from the difference of the oscillator frequency at transmitter and receiver sides and Doppler effect caused by mobility of mobile terminals.
Thus, receivers inevitably experience the frequency offset and needs to estimate the amount of frequency offset and compensate it.
As time goes by, the frequency offset impacts to phases of received symbols more critically, so small amount of the frequency offset can degrade decoder performance and decrease the reliability of whole receiving process. 
Considering that recent communication systems are required to operate moderately in the lower SNR environment, this frequency offset problem needs to be dealt seriously for enhancing the performance in the lower SNR environment.

The frequency offset issue has been dealt for past several decades and various schemes for estimating and compensating frequency offset were proposed through research works. 
A typical approach of estimating frequency offset in Orthogonal Frequency Division Multiplexing(OFDM) systems is to use Cyclic Prefix(CP), which exploits the property of CP that it repeats the tail of an OFDM symbol. (\cite{vandebeek} - \cite{cheng}) 
Assuming that wireless channel is static during the duration of an OFDM symbol, a receiver can estimate the frequency offset from the angular velocity of the received samples in an OFDM symbol.
Since the CP generates the same sample pattern in an OFDM symbol, the receiver can estimate frequency offset by calculating the phase difference of the two samples.
\cite{vandebeek} proposed a joint estimation of timing and carrier - frequency offset by inducing the joint likelihood function of the offset variables.  
\cite{cheng} proposed to exploit the last sample of the cyclic prefix to estimate residual frequency offset, since it lacks interference. 


Another approach measures how the phases of received samples of the known sequence is rotated in time.
This approach can effectively suppress noise variance by using long sequence.
Representative schemes uses preamble sequence which is provided for synchronization purpose by the system. (\cite{naoki}-\cite{schmidl}) 
The Primary Synchronization Signal(PSS)-based method in \cite{naoki}  correlates original PSS sequence and each of the halves of received signal to calculate the phase rotation and estimate frequency offset by dividing the time interval.
\cite{tuf} showed the superior properties of PN-sequence based frequency offset estimator with theoretical analysis. 
\cite{xiao} proposed a two-step method that estimates frequency offset in large and precisely using pseudo noise sequence (PN-sequence).
\cite{schmidl} also proposed two-step method to estimate integer frequency offset with even PN-sequence initially and then to estimate fractional frequency offset with full PN-sequence.  

Some research works approach to estimate CFO from channel gain which can be obtained by pilot symbols lied on frequency domain.
This approach has a potential to estimate CFO more accurately, because the channel gain estimated on frequency domain is free from interference between symbols.
\cite{zeng} uses pilot symbols to estimate channel and compares calculate-phase-then-average method with average-first-then-calculate method.
\cite{boai} proposed the estimator of the integer frequency offset by correlating pilots with other pilots.
In addition, \cite{tanda} proposed a blind estimator by using minimum mean-squared error estimators.

While compensating frequency offset is rather straight-forward, estimating frequency offset is somehow a difficult work and requires more sophisticated algorithms.
When estimating frequency offset, we fundamentally meet several difficulties in a practice.
Most of the frequency offset estimators utilize received signal and the noise contained in the received signal makes the estimator difficult to estimate the precise value of the frequency offset. (This will become more severe when SNR is low.)
Also, As we can observe from the conventional research works, frequency offset is basically estimated from the phase difference measured at two distinct timing.
The frequency offset can only be observed from the angular velocity, so the frequency offset estimator basically requires to calculate symbol phases at two different moments.
This requires more complex mechanism compared to other estimations which can be finished in one-shot.
When designing the frequency offset estimator, it is therefore requires to consider rather complex and detail algorithm and procedure.

In order to perform more detailed frequency offset estimation, we need to further consider the characteristics of the target communication system.
In a practice, a pilot-based estimator generally outperforms  a data-aided estimator and this pilot-based estimator relies on a frame structure of the communication system.
In addition, each communication system has a requirement of target Bit Error Rate (BER) at a specific SNR environment and this requirement heavily depends on the accuracy of the frequency offset estimator.
Otherwise, residual frequency offset due to insufficient accuracy of the estimator will cause unexpected bit errors.
The frequency offset tends to be inaccurately estimated when noise power is strong, so frequency offset estimator needs to be more sophisticated when the system requires more accurate frequency offset estimation.
Thus, the design of the frequency offset estimator needs to be customized according to the system requirements and frame structure.

In this paper, we provide a design methodology for frequency offset estimation for a specific target communication system. 
We start analyzing the estimation accuracy of the frequency offset in a certain level of noise variance.
The result of this numerical analysis will guide the detailed parameters for designing the frequency offset estimator.
Based on the analysis, we design frequency offset estimator for LTE Physical Broadcast CHannel(PBCH), which is relevant to cell search procedure and needs to be decoded in a lower SNR environment.
Based on OpenAirInterface code, we implement our estimator in a software modem form and evaluate our design in a real-time environment by decoding PBCHs from commercial eNodeBs.

\begin{figure}
\includegraphics[width=0.5\textwidth]{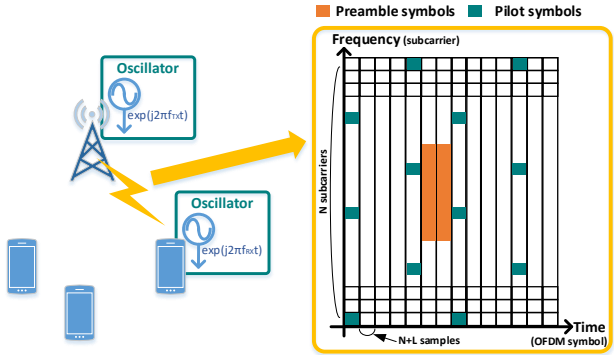}
\caption{\label{fig:frog}System model and frame structure}
\label{model}
\end{figure}

\section{The System Models and Assumptions}

As shown in Fig.~\ref{model}, we assume a general wireless system which composes of a base station and multiple terminals.
We focus on a downlink scenario, where the base station is transmitting data to one of the terminals.
Due to oscillator's inaccuracy and Doppler effect, the carrier frequency of the transmitted signal is different with the receiver side carrier frequency by $f_e$.
In practice, each of the terminals plays a role of figuring out and compensating the frequency offset during receiving process.
The terminal needs to have its own frequency offset estimation scheme based on the frame structure of the system.

We assume that the system modulates data based on OFDM and the frame structure is expressed in two-dimensional spaces as illustrated in Fig.~\ref{model}.
An OFDM symbol consists of $N$ time-domain samples which are generated from Inverse Fast Fourier Transform(IFFT) process by $N$ frequency-domain symbols.
Denoting the time-domain sample as $r[n]$, the terminal can obtain the received symbols by transforming received time-domain samples into frequency domain symbols through Fast Fourier Transform(FFT). 
The received symbol on the frequency domain, denoted by $R$, can be expressed as,
\begin{eqnarray}
& R = \exp(-j2\pi f_et +\theta)S + n, 
\end{eqnarray}
where $S$ is the transmitted symbol, $\theta$ is phase rotation by channel gain and $n$ is Additive White Gaussian Noise(AWGN).

\subsection{Conventional frequency offset estimators}
The conventional schemes mostly estimate frequency offset by measuring the phase difference between frequency-domain symbols or time-domain samples taken at two distinct moments. 
The schemes basically pick two frequency-domain symbols or time-domain samples which are supposed to be received in the same phase without frequency offset.
Assuming that channel gain is static, the phase difference between those symbols or samples are then contributed by the frequency offset.
The terminal can then estimate the frequency offset by calculating the phase difference.
This comes out from simply taking correlation between the two symbols or samples and getting the angle of the correlation result.

The frame structure of the system generally includes several parts that the terminal can refer for its frequency offset estimation.
The terminal can evaluate the frequency offset from estimating the speed of phase rotation for a certain static symbol is rotating.
The terminal can basically use preamble and pilot symbols, which is periodically transmitted and whose patterns and frame positions are known to terminals.

\subsubsection{The preamble-based estimator}
The estimator in  \cite{shoujun} uses the symmetric characteristics of the time-domain preamble samples in LTE and 5G, so the phase of the correlation result over the front and back halves is constant.
By taking the correlation of the forward and backward parts of the preamble samples and calculating the phase difference, which is denoted as $\hat{\theta}_{\Delta t}$ we can estimate the angular velocity.
This is expressed is as follows,
\begin{eqnarray}
2\pi \hat{f_e} = \frac{\hat{\theta}_{\Delta t}}{\Delta t} = \angle \bigg(\bigg[\sum_{k=0}^{N/2-1} r[k] p^*[k]\bigg]^*\cdot\bigg[\sum_{k=N/2}^{N-1} r[k] p^*[k]\bigg]\bigg)
\end{eqnarray}
where $\angle(\cdot)$ indicates the phase and $p[k]$ is time-domain samples transformed from the preamble sequence which is known to terminals.

\subsubsection{The pilot-based estimator}

The terminal can also estimate the angular velocity from the change of the frequency response of the channel, denoted as $H$, and this can be calculated from the received pilot symbols.
Considering the time interval between consecutive pilots is less than a coherent time and the channel is flat-fading, $H$ is estimated to be static for the consecutive pilots.
\cite{zeng} uses this characteristics and evaluates the angular velocity by calculating phase change of $H$.
Denoting $X_1$ and $X_2$ by the consecutive pilot symbols allocated to the same subcarrier and $Y_1$ and $Y_1$ by the received pilot symbols. 
The $H$ at the moments of receiving $Y_1$ and $Y_2$ are evaluated as
\begin{eqnarray}
H_1 = \frac{Y_1}{X_1} \, , H_2 = \frac{Y_2}{X_2}.
\end{eqnarray}
The phase change can be evaluated by summation of phase change over subcarriers. 
This can be calculated as following,
\begin{eqnarray}
2\pi \hat{f_e} = \frac{\angle \sum_{i=1}^{N} H_1^* H_2}{\Delta t}.
\end{eqnarray}

\subsubsection{The cp-based estimator}

As well as the preamble and pilot symbols, the terminal can also utilize CP for the frequency estimation.
We assume that the base station inserts CP with size $L$ by repeating the tail samples. 
From the time-domain signal of size $N$, an OFDM symbol with size $N+L$ is constructed by copying the $L$ tail samples.
The terminal can expect that the initial $L$ time domain samples are equal to the last $L$ time domain samples within an OFDM symbol.
\cite{vandebeek} correlates the time-domain samples of the CP and in the tail part of the OFDM symbol, as follows, 
\begin{eqnarray}
2\pi \hat{f_e} = \frac{\sum_{k=1}^{L} r(k) r^*(k+N)}{\Delta t}
\end{eqnarray}

\subsection{Problem statement}
The ultimate goal is to provide an overall design methodology for robust CFO compensation schemes which essentially include CFO estimation procedures suitable to a specific system frame structure.
The designed scheme eventually estimates CFO $\hat{f_e}$ in a sufficiently precise level so that the system achieves the target level of decoding success probability in a certain SNR environment.
Assuming the system requires the target decoding error probability as $P_e$, we aim to design the overall scheme that can eventually and constantly make $f_e$ less than a certain level with the probability of $1-P_e$.
The robustness of the designed scheme depends on the variance of residual CFO $f_e-\hat{f_e}$ after the scheme is applied, and the require a numerical analysis of $\hat{f_e}$ for a given CFO estimator in a probabilistic perspective. 
The probability distribution of $\hat{f_e}$ can give us an idea of the probability of decoding error in a given SNR environment and we can check whether the corresponding CFO estimator causes decoding error with the probability less than $P_e$ or not.

To see if the design methodology is valid, we provide an example of designing a CFO compensation scheme which is suitable to LTE PBCH decoding.
The LTE PBCH decoding is a good target for evaluating the performance of CFO compensation since the LTE requires that the PBCH should be decoded in low SNR environments and channel estimation by Cell-specific Reference Signal(CRS) during the PBCH decoding process is sensitive to CFO.
(5G PBCH decoding is less dependent to CFO since pilot is allocated for every OFDM symbol in the 5G PBCH.)
From observation of the LTE frame structure, we evaluate how each candidate CFO estimator can provide $f_e$ accurately.
The residual CFO generated by the designed scheme should be lower than a threshold $f_{e,max}$ that causes decoding error with the probability of $P_e$. ($f_{e,max}$ and $P_e$ will be derived for the LTE system.)

\section{Numerical Analysis for CFO Estimation Error}
Estimation error is a significant factor of evaluating how an estimator is robust 
To see how precise a specific CFO estimator provide the estimated CFO $\hat{f_e}$, it is needed to derive the probability distribution of $\hat{f_e}$.
There are several factors that affect to the probability distribution of $\hat{f_e}$ in the CFO estimator.
One important factor is the time interval $\Delta t$ during which the CFO estimator calculates angular velocity.
We can commonly find from the above CFO estimators that the amount of phase change is calculated at two moments and $f_e$ is induced by normalizing $\Delta t$.
Also, each CFO estimator evaluates phase difference in a form of averaging process for obtaining reliable angular velocity in a noisy environment.
Considering the effect of noise variance reduction, we denote $\sigma_{n,est}^2$ as the variance of noise after the estimator's process.
We assume that the estimator doesn't not guarantee to obtain accurate provide due to this noise and $\hat{f_e}$ essentially contains an error $\Delta f_e$.

\subsection{The Generalized Probability Model for a CFO Estimator}

Since a certain level of frequency offset directly causes to the decoding failure, it is needed to reduce the maximal frequency offset error $\Delta f_{e,max}$ for meeting the criterion of the maximum error rate $P_e$.
If the base station transmits a symbol $s$ with symbol energy $E_S$ at the two moments of interval $\Delta t$, received symbols at the two moments $r_1$ and $r_2$ will be as following
\begin{eqnarray}
r_i &= e^{j \theta_i} s + n_i=e^{j \hat{\theta_i}}, i=1,2
\end{eqnarray}
where $\theta_1$ and $\theta_2$ are the phase rotations by frequency offset and channel. 
Here, $n_1$ and $n_2$ are the effective noise terms which are the output of an frequency offset estimator of AWGN and whose variance is $\sigma_{n,est}^2$.
To see how the noise impacts in a phase perspective, the effective noise terms are expressed in a complex form as following, 
\begin{eqnarray}
n_i = n_{ix} + j n_{iy} , \, i = 1,2,
\end{eqnarray}
where $j=\sqrt{-1}$. $n_{ix}$ and $n_{iy}$ are normal random variables whose variance is $\frac{\sigma_{n,est}^2}{2}$. 
To derives $f_e$, an frequency offset estimator $f$ is set to obtain ${\theta_2 - \theta_1}$ as following,
\begin{eqnarray}
f(r_1, r_2) = \hat{\theta_2}-\hat{\theta_1} = \angle(r_1^* r_2)
\end{eqnarray}

\subsection{The Probability Distribution of the Phase Difference Estimation}

\begin{figure} [h]
\includegraphics[width=0.5\textwidth]{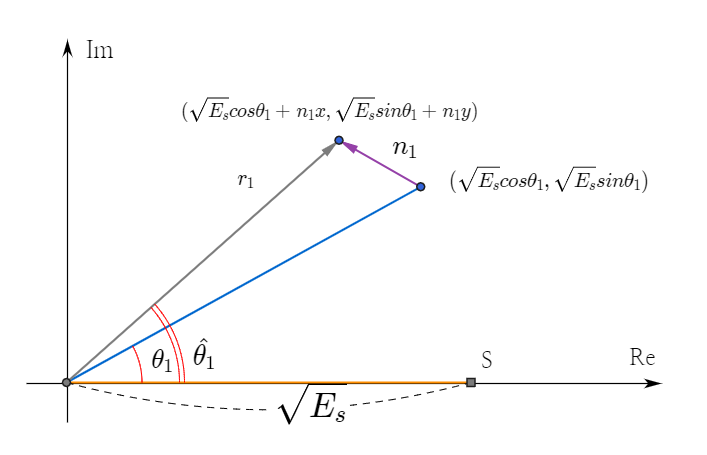}
\caption{\label{fig:frog}Signal space diagram of received symbol and the effective noise}
\label{symbol}
\end{figure}

One way to derive the probability distribution of $\hat{\theta_2} - \hat{\theta_1}$ is to see the tangent.
This is valid when frequency offset compensation is successfully performed so that the residual frequency offset is nearly zero and $\tan(\hat{\theta_2} - \hat{\theta_1}) \approx \hat{\theta_2} - \hat{\theta_1}$.
Using the tangent subtraction formula, $\tan(\hat{\theta_2} - \hat{\theta_1})$ is re-expressed with $tan(\hat{\theta_i})$ as following
\begin{eqnarray}
\tan{(\hat{\theta_2}-\hat{\theta_1})} = \frac{\tan\hat{\theta_2}-\tan\hat{\theta_1}}{1+\tan{\hat{\theta_2}}\tan{\hat{\theta_1}}}. \label{phasediff}
\end{eqnarray}
$\hat{\theta_i}$ which is randomized by noise terms $n_{ix}$ and $n_{iy}$ as illustrated in Fig.~\ref{symbol}.
Based on the received symbol model, the tangents of the received symbols' phase can be derived as following,
\begin{eqnarray}
 \tan{\hat{\theta_i}}=\frac{\sqrt{E_s} \sin{\theta_i}+n_{iy}}{\sqrt{E_s} \cos{\theta_i}+n_{ix}}, i=1,2. \label{tangent}
\end{eqnarray}

By substituting (\ref{tangent}), (\ref{phasediff}) is expressed as following,
\begin{eqnarray}
&\tan{(\hat{\theta_2}-\hat{\theta_1})}=\frac{\frac{\sqrt{E_s} \sin{\theta_2}+n_2 y}{\sqrt{E_s} \cos{\theta_2}+n_2 x}-\frac{\sqrt{E_s} \sin{\theta_1}+n_{1y}}{\sqrt{E_s} \cos{\theta_1}+n_{1x}}}{1+\frac{\sqrt{E_s} \sin{\theta_2}+n_2 y}{\sqrt{E_s} \cos{\theta_2}+n_{2x}} \frac{\sqrt{E_s} \sin{\theta_1}+n_{1y}}{\sqrt{E_s} \cos{\theta_1}+n_{1x}}}.\\
&=\frac{E_s \sin{(\theta_2 \! - \! \theta_1)} \! + \! \sqrt{E_s}X_1 \! + X_2}{E_s \cos{(\theta_2 \! - \! \theta_1)} \! + \! \sqrt{E_s}Y_1 \! + Y_2}, \label{tangentFrac}
\end{eqnarray}
where
\begin{eqnarray}
& X_1 = \cos{\theta_1} n_{2y} + \sin{\theta_2} n_{1x} -\cos{\theta_2} n_{1y} - \sin{\theta_1} n_{2x} \\
& Y_1 = \cos{\theta_2} n_{1x} + \cos{\theta_1} n_{2x} +\sin{\theta_2} n_{1y} + \sin{\theta_1} n_{2y}\\
& X_2 = n_{2y} n_{1x} - n_{1y} n_{2x}\\
& Y_2 = n_{2x} n_{1x} - n_{2y} n_{1y}.
\end{eqnarray}
Since the effective noise terms $n_{ix}$ and $n_{iy}$ are independent, $X_1$ and $Y_1$ follow normal distribution whose mean and variance are $0$ and $\sigma_{n,est}^2$, respectively.
In addition, $X_2$ and $Y_2$ are equivalent to multiple of two independent normal random variables whose mean and variance are $0$ and $\frac{\sigma_{n,est}^4}{2}$.

Thus, the numerator and denominator parts of (\ref{tangentFrac}) have similar probability distributions whose mean are $\sin(\theta_2 - \theta_1)$ and $\cos(\theta_2 - \theta_1)$, respectively, and variance is $E_s \sigma_n^2 + \frac{\sigma_n^4}{2}$.
By referring the distribution of the ratio of two variables from \cite{diaz}, the mean of (\ref{tangentFrac}) is the ratio of means of the numerator and denominator, which is eventually expressed as
\begin{eqnarray}
E[\tan{(\hat{\theta_2}-\hat{\theta_1})}] = \tan{(\theta_2-\theta_1)} \approx \theta_2-\theta_1,
\end{eqnarray}
and the variance can be derived as
\begin{eqnarray}
& Var[\tan{(\hat{\theta_2}-\hat{\theta_1})}]\\
& = \tan^2{(\theta_2-\theta_1)} \bigg(\frac{E_s \sigma_n^2 + \frac{\sigma_n^4}{2 }}{E_s^2 \sin^2{(\theta_2 - \theta_1)}} +  \frac{E_s \sigma_n^2 + \frac{\sigma_n^4}{2}}{E_s^2 \cos^2{(\theta_2 - \theta_1)}}\bigg) \\
& = \bigg( \frac{ \sigma_{n}^{2} }{E_s} + \frac{1}{2} \bigg( \frac{\sigma_n^2}{E_s} \bigg) ^2 \bigg) \frac{1}{\cos^4{(\theta_2 - \theta_1)}} 
\end{eqnarray}

Here, $\frac{\sigma_n^2}{E_s}$ is an effective Signal to Noise Ratio(SNR) which is the SNR after the frequency offset estimator. We denote the effective SNR by $SNR_{e}$. Since  
$\hat{\theta_2}-\hat{\theta_1} = 2 \pi \hat{f_e} \Delta t$
the variance of $\hat{f_e}$ can be derived as
\begin{eqnarray}
 V[\hat{f_e}]=\bigg(\frac{1}{2 \pi \Delta t}\bigg)^2\bigg(\frac{1}{2 SNR_e^2} + \frac{1}{ SNR_e}\bigg) \frac{1}{\cos^4{(2 \pi f_e \Delta t)}}.\label{resVar}
\end{eqnarray}

(\ref{resVar}) reveals how much the estimated CFO $\hat{f_e}$ deviates with respective to the parameters of the CFO estimator and the signal environment. 
The result of the CFO estimator varies largely if time interval between the two phase measurements is small or the effective SNR is low.
This means that for any frequency offset estimator, the result of frequency offset estimation tends to swing small if the received signal strength is large or the estimator effectively reduces noise variance.
Also we can learn from it that the frequency offset estimator stably provides more accurate values and if the time interval between the measurements is set to be large. 
It is noted that $f_e$ converges to small as frequency offset compensation is performed successively and we can assume that $\cos(2\pi f_e \Delta t) \approx 1$.
So, the term $\frac{1}{\cos^4(\theta_2 - \theta_1)}$ in (\ref{resVar}) impacts less to the overall variance of the frequency offset estimation results.

\subsection{Criterion for Satisfying Target Decoding Performance}
We simply assume that decoding error occurs when the frequency offset becomes larger than a threshold $\Delta f_{e,max}$.
Then decoding error probability is equal to the probability that the error of frequency offset estimation is larger than $\Delta f_{e,max}$. In detail,
\begin{eqnarray}
P(|\hat{f_e}-f_e| > \Delta f_{e,max}) = P(\hat{f_e} > f_e + \Delta f_{e,max}) \leq \frac{P_e}{2}. \label{condition}
\end{eqnarray}
Based on (\ref{resVar}), (\ref{condition}) holds if $\Delta f_{e,max}$ satisfies following inequality;
\begin{eqnarray}
 \Delta f_{e,max} \geq Q^{-1}\bigg(\frac{P_e}{2}\bigg) \sqrt{\bigg(\frac{1}{2 \pi \Delta t}\bigg)^2\bigg(\frac{1}{2 SNR_e^2} + \frac{1}{ SNR_e}\bigg)}.\label{criterion}
\end{eqnarray} 

For designing frequency offset estimation, (\ref{criterion}) can be used as a criterion for satisfying the system target.
For a given value $\Delta f_{e,max}$, (\ref{criterion}) can be satisfied if the frequency offset estimator is designed to measure during long time interval and have higher $\Delta t$.
Also, the frequency offset estimator is likely to satisfy (\ref{criterion}) if it reduces noise variance effectively and have higher $SNR_e$.
In addition, it is trivial that the inequality is likely to be satisfied if the target decoding error probability $P_e$ is high or a decoder used in the system performs better so that $\Delta f_{e,max}$ is large.
It should be noted that one is advantageous to choose $\Delta t$ as a large value in view of (\ref{criterion}) and also is advantageous to choose it as a small value in view of estimation range.
If the estimator has too small $\Delta t$, then it cannot estimate sufficiently large frequency offset.

\begin{figure} [h]
\includegraphics[width=0.5\textwidth]{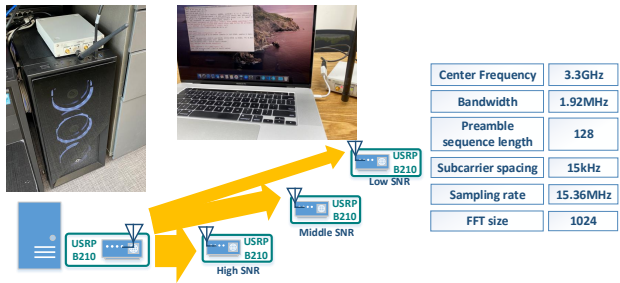}
\caption{\label{fig:frog} Experiment Environments for Measuring CFO estimation}
\label{tb1}
\end{figure}

\subsection{Comparison with measurement results of frequency Offset estimation}
For evaluating the numerical model of the frequency offset estimation, we conducted an experiment with a real-time system and compares our numerical model with the histogram of the measurement results of the real frequency offset.
As shown in Fig.~\ref{tb1}, we used two USRP B210s with 3.5GHz antennas and a static PC and a notebook for baseband processing.
The detailed parameters used in the experiment is described in the right-hand side of the Fig.~\ref{tb1}.
We tested for the preamble-based CFO estimator mentioned in Section 2, where $\Delta t = 33.3us$. 
We implemented simple baseband process software that controls the transmitter side to generate preamble signal by using M-sequence as preamble sequence and the receiver side to estimate CFO values repeatedly by the above estimator with real signal samples.

\begin{figure} [h]
\includegraphics[width=0.5\textwidth]{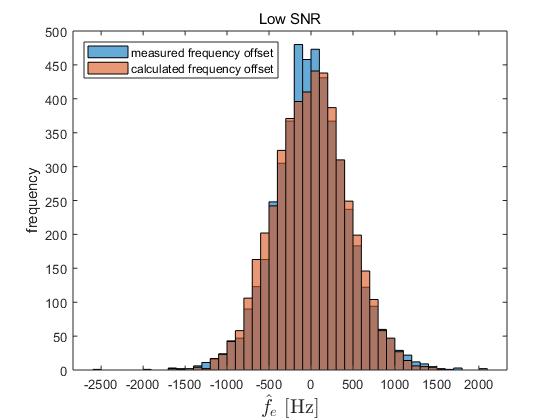}
\caption{\label{fig:frog} Numerical and measurement results of CFO estimation when $SNR = 5.6434dB$}
\label{model_low}
\end{figure}

\begin{figure} [h]
\includegraphics[width=0.5\textwidth]{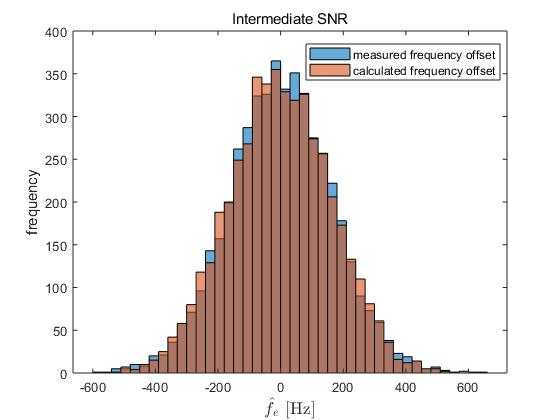}
\caption{\label{fig:frog} Numerical and measurement results of CFO estimation when $SNR = 14.2635dB$}
\label{model_mid}
\end{figure}

\begin{figure} [h]
\includegraphics[width=0.5\textwidth]{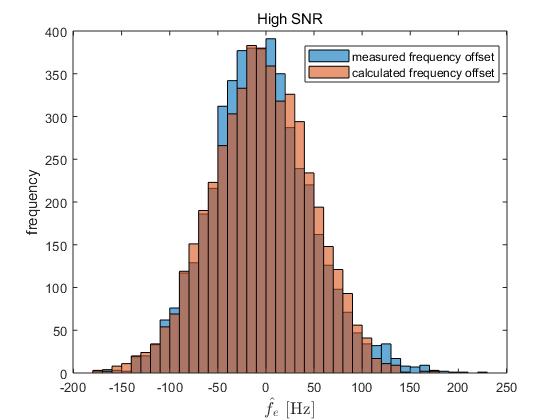}
\caption{\label{fig:frog} Numerical and measurement results of CFO estimation when $SNR = 24.2859dB$}
\label{model_high}
\end{figure}

We have collected estimated CFO values with allocating the receiver side system to three different positions for making various SNR environments.
Fig.~\ref{model_low}-\ref{model_high} show the histograms of the frequency offset derived from our numerical model and the measured frequency offset in case of three different SNR environments. 
We can verify from the figures that the numerical model fits to the measurement results for various SNR conditions.
This proves that our numerical analysis for frequency offset estimation is valid and can be used for evaluating the probability of frequency offset estimations.
As we can find from the numerical model, frequency offset estimation varies greatly as SNR becomes lower.
This indicates that more detailed frequency offset estimator should be applied in lower SNR environment if the system needs to estimate frequency offset accurately and keep the decoding performance in a high-level for low SNR environment.
This can be achieved by using an estimator with measuring phases for large time interval $\Delta t$ or averaging greatly to achieve low $\sigma^2_{n,est}$.

\section{Design Example of Frequency Offset Estimation Scheme for LTE PBCH Decoding}
Based on the above criterion, we can design an overall scheme of CFO estimation and compensation that is customized to a specific system with meeting the target decoding performance.
The condition (\ref{criterion}) reveals whether a specific CFO estimator causes significant CFO error and keeps away from meeting the system's target decoding performance.
In other words, we can use the condition (\ref{criterion}) to check how a CFO estimator can perform in terms of decoding performance with given system parameters $P_e$ and $\Delta f_{e,max}$ when we design an overall CFO estimation scheme of the system.
This is 
As an example, we here design the CFO estimation and compensation scheme for LTE Physical Broadcast CHannel(PBCH) decoding by using the condition (\ref{criterion}).
We set the target system as LTE (instead of 5G) since the LTE is more challenging in view of the CFO issue and its performance depends more critically on how the CFO issue is dealt.

\begin{figure} [h]
\includegraphics[width=0.5\textwidth]{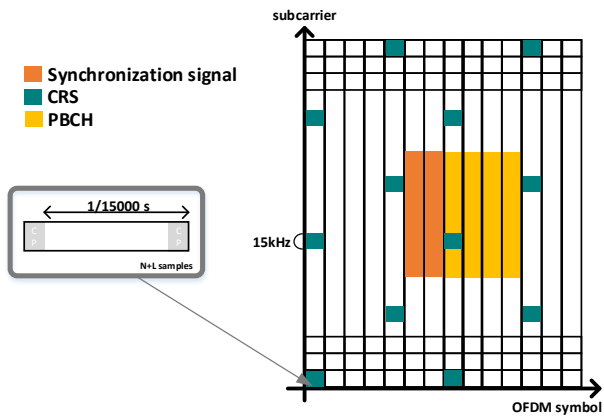}
\caption{\label{fig:frog} Frame structure of LTE}
\label{LTEPBCH}
\end{figure}

\subsection{Analysis of the LTE PBCH decoding procedure}
Fig~\ref{LTEPBCH} illustrates the frame structure of the LTE.
The subcarrier spacing of the LTE is $15kHz$, so the duration of the OFDM symbol without CP is $1/15000$.
The transport channel codes the system information bits by convolutional encoder with repetition and eventually generates 1920 bits. 
These bits are modulated by Quadrature Phase Shift Keying(QPSK) and allocated to the PBCH region that occupies 4 OFDM symbols.
For channel estimation, the LTE constantly transmits pilot symbols called Cell specific Reference Signal(CRS), and the time interval between the neighboring CRSs near the PBCH region is 4 OFDM symbols.

Before designing the CFO estimation scheme, we firstly needs to define $P_e$ and $\Delta f_{e,max}$.
We can define $P_e$ as the target detection probability of PBCH in the perspective of cell search or target BLock Error Rate(BLER) so we can assume $P_e = 10\%$ in case of LTE PBCH decoding.
To derive $\Delta f_{e,max}$ for the LTE PBCH decoding, let us assume that decode error happens if the symbol phase is distorted within the duration of PBCH transmission.
Since PBCH uses QPSK, the phase difference of the neighboring symbols is $\pi/2$.
Considering maximum likelihood detection, the received symbol is mis-detected if the phase rotation due to frequency offset exceeds $\pm \pi/4$ within $4$ OFDM symbols.
Since the subcarrier spacing in LTE is $15kHz$, the above angular velocity corresponds to $\Delta f_{e,max} = \frac{15000}{2\cdot 4 \cdot 8}=234.4Hz$.
In practical, symbol detection error can be recovered by Viterbi decoder so $\Delta f_{e,max}$  can be assume as larger than above derivation, so we will simply assume that $\Delta f_{e,max} = 300Hz$ for the case of LTE PBCH decoding.

Since the frame structure of the LTE contains CP and CRS whose time intervals are different, we can apply both CP-based and CRS-based CFO estimators that have distinct characteristics.
The CP-based estimator will provide CFO values {\it coarsely}, since it evaluate phase difference whose time interval is an OFDM symbol duration and $\Delta t$ is relatively short.
The CRS-based estimator, on the other hands, will provide more detailed CFO values, since the time interval of consecutive CRSs is relatively long compared to the OFDM symbol duration.
Thus, we can consider to utilize the CRS-based estimator for compensating {\it residual} CFO.

For designing the overall procedure of CFO estimation and compensation that is suitable for the LTE PBCH decoder, it is needed to check the criterion for the candidate estimators in a certain SNR environment.
The right hand in (\ref{criterion}) corresponds to the maximal frequency offset estimation error with probability $P_e$ when the effective SNR is $SNR_e$.
With given values of $\Delta f_{e,max}$ and $P_e$, we can see the minimum SNR that the CP-based and CRS-based frequency offset estimators can satisfy the criterion.
This leads to select a proper frequency offset estimator which can satisfies the performance requirement of LTE PBCH decoder.
As well as the decoder performance, we also need to consider the range of frequency offset that each candidate estimator can evaluate.
In case of an estimator with large $\Delta t$, the phase difference $\theta_2 - \theta_1 = 2\pi f_e \Delta t$ will exceed $2\pi$ and frequency offset will wrongly estimated.
The estimator can only estimator frequency offset with range $[-\frac{f_s}{2\Delta t} ~ -\frac{f_s}{2\Delta t}]$.

\begin{figure}
\includegraphics[width=0.5\textwidth]{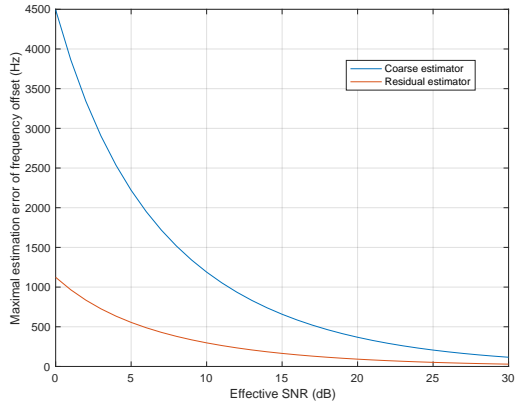}
\caption{\label{fig:frog}Comparison of frequency offset estimation error for coarse and residual estimator in case that $P_e = 10\%$}
\label{mse}
\end{figure}

In case of CP-based estimator, which we call here coarse estimator, $\Delta t$ is 71.429us, which is relatively short. 
This characteristics enables the coarse estimator to evaluate large frequency offsets, since the range of frequency offset estimation is $[-7002.8 ~ 7002.8]Hz$.
However, the coarse estimator is not good for figuring out the frequency offset accurately, since the maximal frequency offset estimation error for the probability 0.1 is rather high.
In Fig.~\ref{mse}, the maximal frequency offset estimation error becomes lower than $\Delta f_{e,max}$ of the LTE PBCH decoder when the effective SNR is larger than $22dB$.
This means that coarse estimator can satisfy the performance requirement of LTE PBCH decoder if the effective SNR is larger than $22dB$, and cannot satisfy it if the SNR is low.

To deal the decoder performance issue, we can also consider a CRS-based estimator, which we call here a residual estimator, since $\Delta t$ is 4 OFDM symbol duration which corresponds to $285.716us$.
The residual estimator can estimate more accurate frequency offset than the coarse estimator, since $\Delta t$ is higher.
In Fig.~\ref{mes}, we can see that in case that effective SNR is $10dB$ the residual estimator can perform frequency offset error lower than $300Hz$.
This means that the residual estimator can perform $12dB$ better than the coarse estimator in the effective SNR perspective.
The residual estimator can however evaluate limited range of frequency offset, so it will mis-operate if the actual frequency offset is large. 
(This can happen in case of initial cell search, where the terminal has not yet compensated frequency offset.)
The residual estimator can estimate frequency offset in the range of $[-1752 ~ 1752]Hz$.

\begin{figure}
\includegraphics[width=0.5\textwidth]{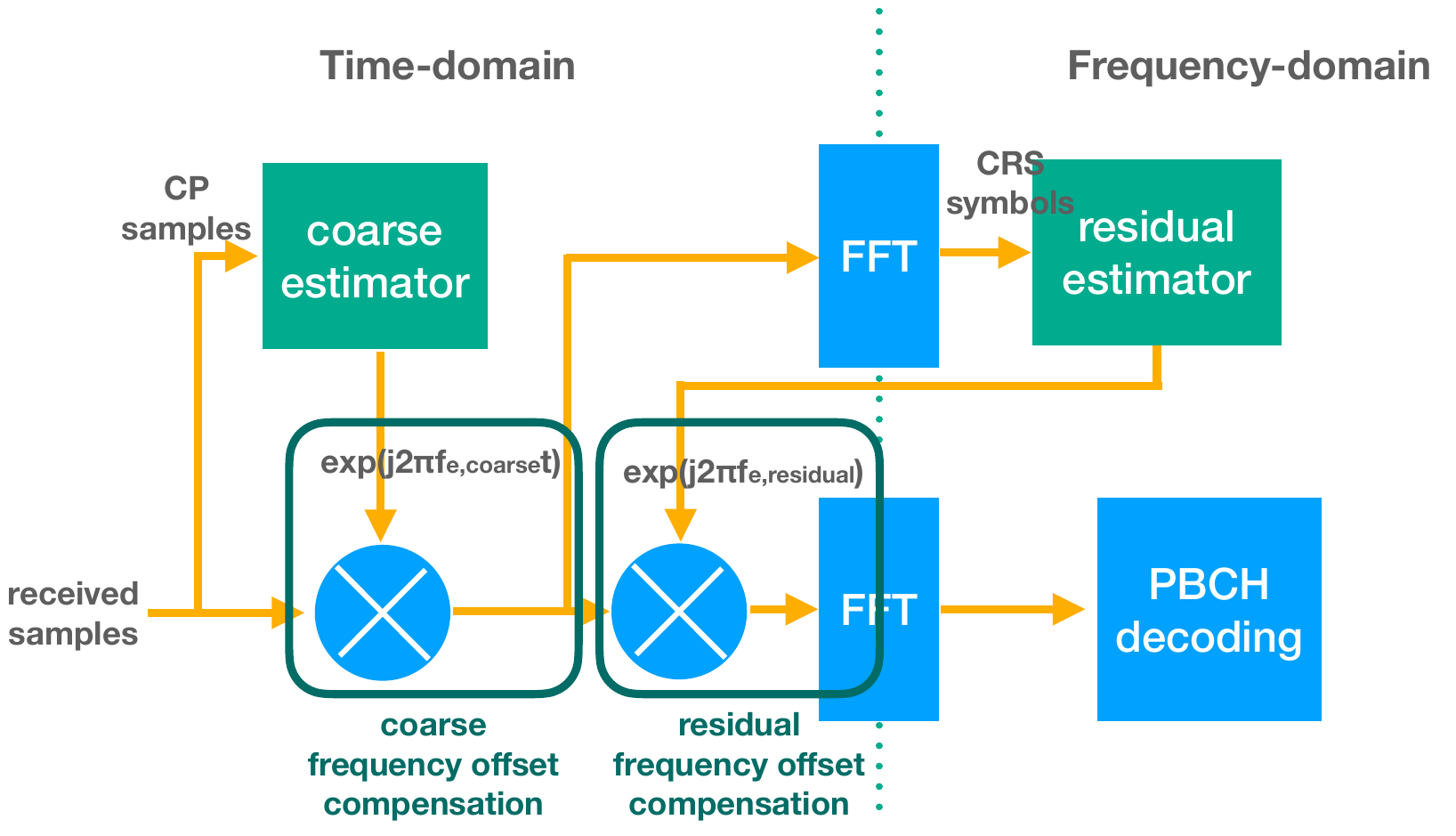}
\caption{\label{fig:frog}Proposed frequency offset estimation scheme using coarse and residual estimators}
\end{figure}

\subsection{Two-step CFO estimation and compensation scheme}
Since the coarse and residual estimators have conflict characteristics, we need to consider a hybrid design which uses both coarse and residual estimators.
Fig.~\ref{twostep} illustrates the overall procedure of the proposed CFO estimation and compensation scheme for the LTE PBCH decoding.
The proposed scheme estimates and immediately compensates CFOs in two steps before frequency-domain processing.
The initial step estimates coarse CFO estimation for compensate relatively large CFO roughly.
It also has the latter step of residual CFO estimation for compensating residual CFO after the coarse CFO compensation.

Table.~\ref{algorithm} shows the detailed operations of the proposed scheme.
The coarse CFO compensation estimates CFO by the CP-based estimator.
This can be conducted at the initial phase of PBCH decoding with using the knowledge of OFDM symbol boundary, since it only requires to know the positions of the CP samples in time-domain.
Once the system recognizes the boundary of OFDM symbols by a synchronization process, the proposed scheme can extract CP and the original post samples from the boundary timing information.
After calculating coarse CFO $f_{e,coarse}$ from (\ref{coarse}), frequency compensation by multiplying $\exp{j2\pi f_{e,coarse}}n$ to each received samples of the 4 OFDM symbols in PBCH region.

The residual CFO compensation estimates CFO by the CRS-based estimator with the knowledge of both time and frequency domain frame structure. 
Since the estimator needs to process symbols on frequency domain, a certain level of frequency domain synchronization should be performed prior to the this step.
(This is achieved by coarse CFO compensation in the proposed scheme.)
The CRS-based estimator also requires to know the pattern of the CRS, which depends on the cell ID, so this can be performed after synchronization signal is identified.
These cause the residual CFO compensation to be done after the coarse CFO compensation and during the frequency-domain processing.
Based on the cell ID, the estimator collects the received symbols that correspond to CRSs and in the OFDM symbols 7 and 11 and measures phase difference.
Using the residual CFO $f_{e,residual}$, frequency compensation by multiplying $\exp{j2\pi f_{e,residual}}n$ to each received samples of the 4 OFDM symbols in PBCH region and re-start the initial step of the frequency-domain processing.

By compensating CFO in the two steps, the proposed scheme can handle CFO more robustly in various CFO and SNR environments.
Since the coarse estimator can handle wide range of CFO, the large-scale CFO can be compensated in prior.
The initial step can also encourage the latter frequency domain processing such as synchronization signal detection and collecting CRS symbols for the residual CFO estimation.
In addition, the residual estimator can make up the inaccuracy of the initial coarse CFO compensation.
This detailed step can enhance the decoding performance when the SNR is low and the decoder is greatly affected by small amount of CFO.
Although the residual estimator can recognize the small range of CFO, the residual estimator can work well since the input of the residual estimator is already compensated once in large-scale.
The residual estimator guarantees to provide precise CFO values unless the error amount at the initial coarse CFO compensation step exceeds the range of the residual estimator, i.e. $|f_e - f_{e, coarse}| > 1750Hz$.



\section{Performance Evaluation}

\begin{figure}
\includegraphics[width=0.5\textwidth]{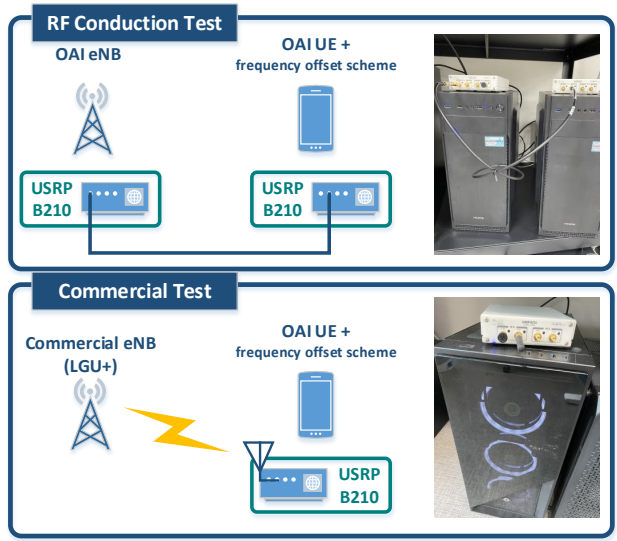}
\caption{\label{fig:frog}Testbed description}
\label{testbed}
\end{figure}
We verify the decoding performance for our frequency offset estimation scheme based on a real-time operating testbed.
Fig.~\ref{testbed} illustrates the testbed configurations.
We realize LTE User Equipment(UE) and eNodeB(eNB) with using software modem systems which are composed of PCs and Radio Frequency (RF) devices.
We use USRP B210s as the RF signal processing and utilize OAI-RAN source code provided by OpenAirInterface as the baseline software of baseband signal processing.\cite{OAI}
In a transmission case, The OAI software running on a PC generates LTE PBCH baseband signal and the USRP B210 generate RF signal based on this baseband signal. 
In a reception case, the OAI software decodes the LTE PBCH baseband signal sampled by USRP B210.

On the OAI source code for LTE UE, we implemented our frequency offset estimation scheme.
Fig.~\ref{codeImp} shows the pseudo-code and the example of the code implementation.
We added 

We also modified the CRS-based estimator code included in the OAI source code.
Since the range of input channel estimation values varies critically, we changed to utilize 64bit variables for calculating inner product.
Consequently we can lower-down the scaling value so that the residual estimator can provide stable CFO values when the amplitudes of received symbols are low.
We also changed the residual CFO is estimated from the OFDM symbol 7 and 11, which are the ones near PBCH, meanwhile the original CRS-based estimator uses the OFDM symbols 0 and 4.

It is noted that original OAI source code includes the CRS-based estimator and we utilized this code for realizing the residual estimator.
(The OAI code doesn't originally conduct coarse frequency offset estimation.)
We completed to implement our proposed scheme by adding the code of coarse estimator.

We have conducted two test scenario; RF conduction and commercial scenarios.
In the RF conduction scenario, the OAI software play a role of LTE eNB and the proposed scheme estimate frequency offset of the eNB signal generated by the OAI software.
We connected the LTE eNB and UE with RF cable, which is equivalent to an AWGN channel.
This scenario is for evaluating the performance of the frequency offset compensation in detailed SNR environments.
In the commercial scenario, 889MHz antenna is attached to the LTE UE and the proposed scheme estimate frequency offset of the wireless signal transmitted from LGU+ commercial eNB.
This scenario is to evaluate how our frequency offset estimation scheme works in practical environments.

\subsection{Performance of LTE PBCH decoder with proposed frequency offset estimation scheme}

\begin{figure} [h]
\includegraphics[width=0.5\textwidth]{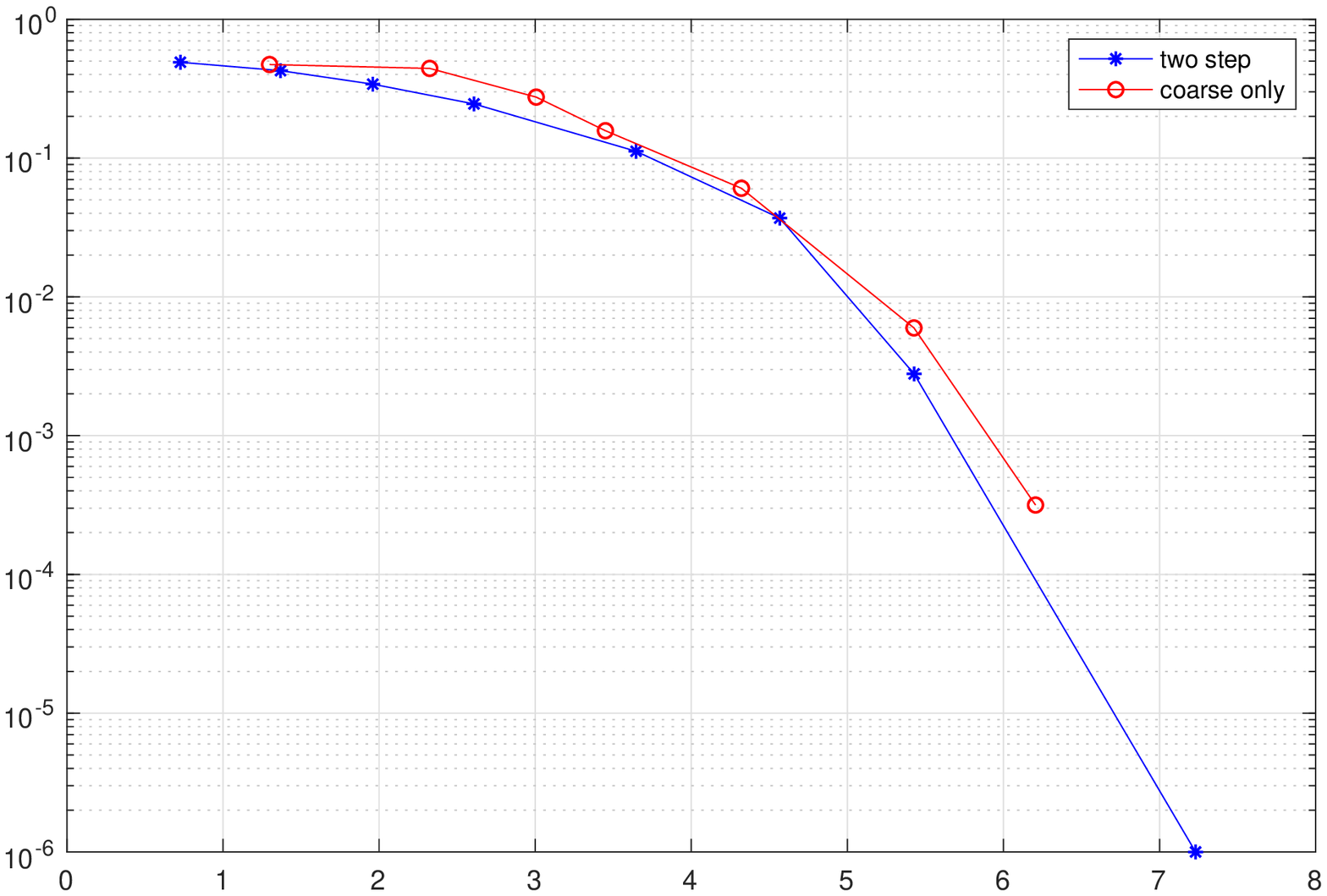}
\caption{\label{fig:frog} }
\label{labTest}
\end{figure}

Fig.~\ref{labTest} show the decoding probability of the LTE PBCH when using coarse estimator, residual estimator and both estimators.
We repeated PBCH decoding and measured the error rate in the condition that the PCI(Physical-layer Cell ID) was detected successfully.
In Fig.~\ref{labTest}, it is shown that as SNR increases, the error rate decreases when the offset is 700Hz. The offset-700Hz-coarse-only graph is always on the right-hand side of the offset-700Hz in the same error rate. Indeed, when the error rate is around $10^{-1}$, there is about an 1dB SNR gain. These results shows why the residual estimation is required. 

On the other hand, in only-residual-compensation case, PBCH decoding barely succeeds. The residual estimation has smaller range of estimated frequency offset, which is not enough to succeed PBCH decoding, which requires large frequency offset compensation. 

\begin{figure} [h]
\includegraphics[width=0.5\textwidth]{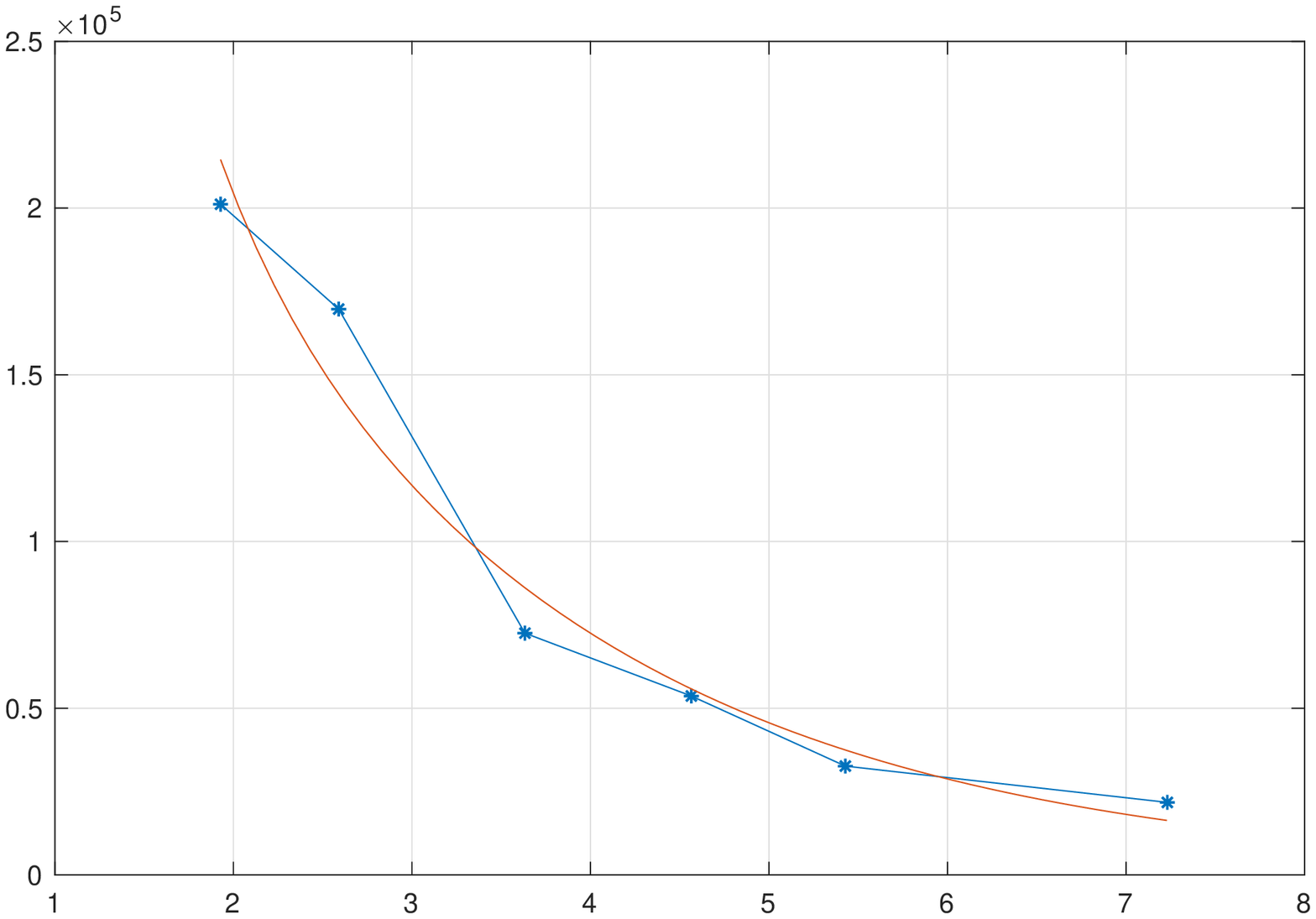}
\caption{\label{fig:frog} Variances of CFO estimation compared with the numerical model}
\label{labTest}
\end{figure}

Besides, we had the other experiment with commercial eNB signals. 
We set two different SNR conditions (high and low), and used two different USRPs as UEs in the high SNR condition. For each three conditions, we applied two-step method, coarse-only method, and residual-only method to measure an success rate of PBCH decoding, only when the USRP detected the Cell ID that the certain eNB sent.  

\begin{figure} [h]
\includegraphics[width=0.5\textwidth]{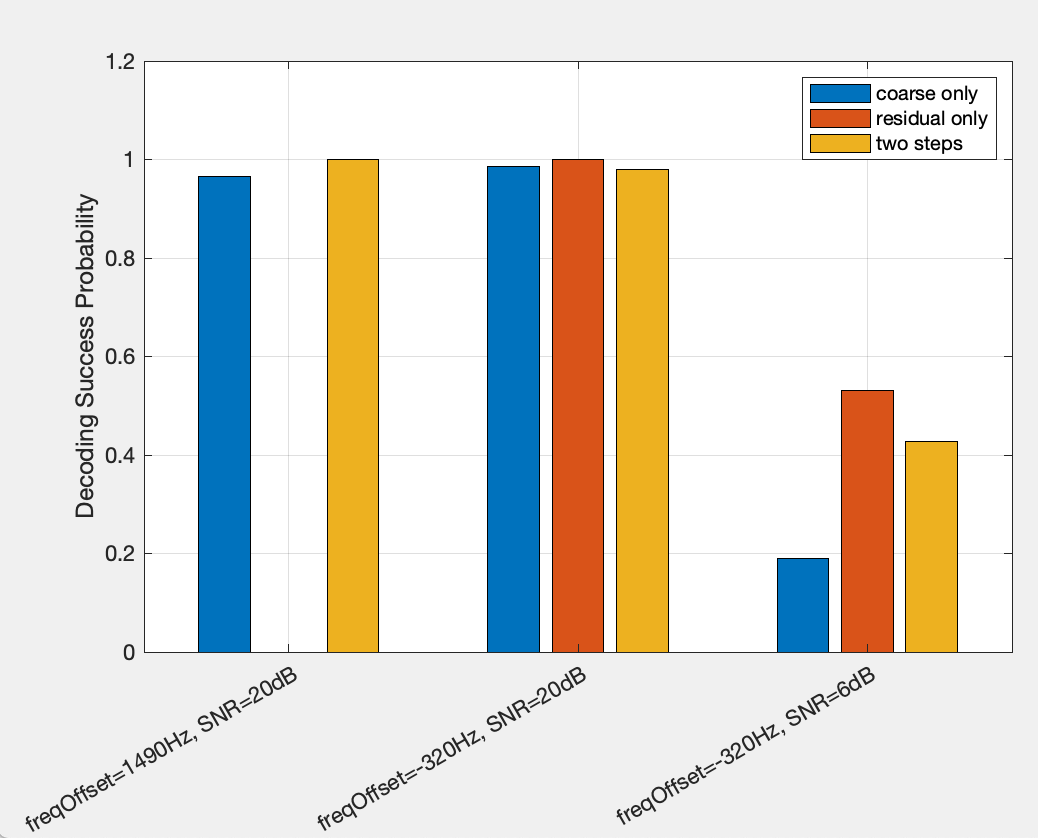}
\caption{ Each condition uses different USRP receivers. Let's name them USRP 1, USRP 2, USRP 3 in order. }
\label{commercialTest}
\end{figure}

As shown in Fig.~\ref{commercialTest}, the frequency offsets of USRP 1 and USRP 2 are different, caused by the difference of the property of the receivers. When frequency offset is small(USRP 2), the PBCH decoding succeeds even with coarse-only estimation, to the same degree as two-step method. However, when the offset is large (USRP 1), the rate of the coarse-only method is lower than that of the two-step method. 

In addition, We can compare USRP 2 and 3, which have similar frequency offset but different SNR. In case of USRP 3, whose SNR is way lower than USRP 2, the two-step method makes the success rate higher than only-coarse method, while URSP 2, whose SNR is high, has no significant difference in succcess rate.



\section{Conclusion}

In this paper, the two-step estimation method is proposed to detect proper frequency offset in a given system. This method includes coarse estimation with small time interval and residual estimation with certain interval that satisfies the maximum frequency offset error to succeed detection and the maximum detection error rate. The experiments we had shows that the distribution we induced mathematically is valid and the proposed method is appropriate for detection.  

%



\section*{Acknowledgment}

This research was supported by Sookmyung Women's University Research Grants (1-1903-2003) and the Institute of Information \& communications Technology Planning \& Evaluation(IITP) grant funded by the Korea government(MSIT) (No. 2021-0-00874, Development of Next Generation Wireless Access Technology Based on Space Time Line Code)

\ifCLASSOPTIONcaptionsoff
  \newpage
\fi



%

%

\begin{IEEEbiography}[{\includegraphics[width=1in,height=1.25in,clip,keepaspectratio]{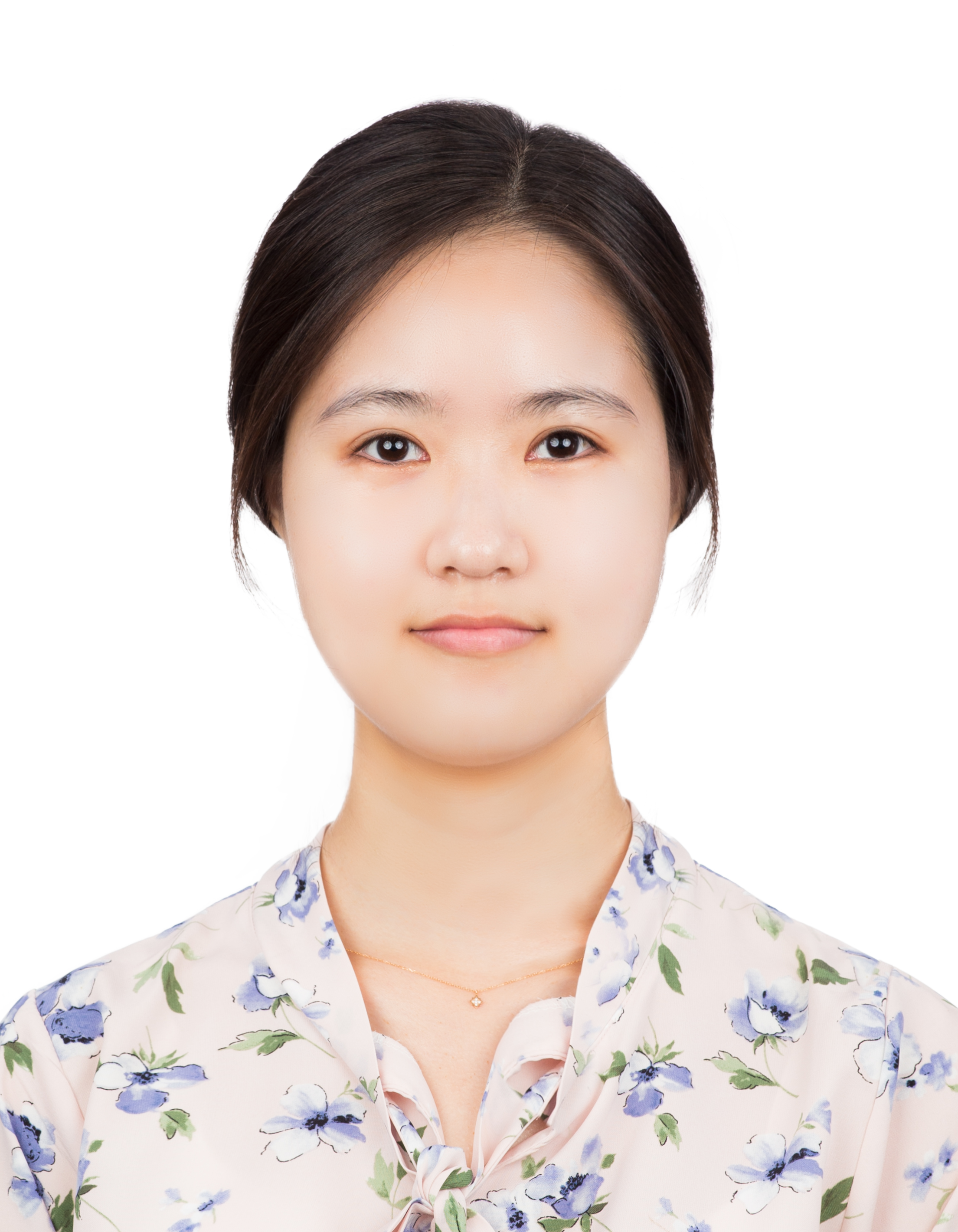}}]{Minkyeong Jeong}
Biography text here.
\end{IEEEbiography}

\begin{IEEEbiographynophoto}{Sang-Won Choi}
Biography text here.
\end{IEEEbiographynophoto}


\begin{IEEEbiography}[{\includegraphics[width=1in,height=1.25in,clip,keepaspectratio]{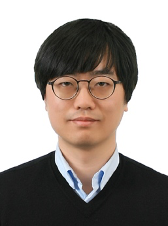}}]{Juyeop Kim} (jykim@sookmyung.ac.kr) is an assistant professor in the Department of Electronics Engineering, Sookmyung Women's University, Seoul, Korea. He received his B.S. and Ph.D. in electrical engineering from Korea Advanced Institute of Science and Technology (KAIST) in 2004 and 2010, respectively. From 2010 to 2011, he was with KAIST Institute IT Convergence Research Center in charge of research for 5G cellular system. From 2011 to 2013, he was with Samsung Electronics in charge of development and commercialization for 2G/3G/4G multi-mode mobile modem solution. From 2014 to 2018, he was with Korea Railroad Research Institute in charge of research and development for LTE-Railway(LTE-R), Public Safety LTE (PS-LTE) and railway IoT solutions. His current research interests is software modems and next-generation wireless communications systems.
\end{IEEEbiography}




\end{document}